\documentclass[footinbib, aps,prl,reprint, 10pt,twocolumn,showpacs, superscriptaddress]{revtex4-2}

\usepackage{amsfonts,amsmath,amssymb}
\usepackage{graphicx}
\usepackage[pdfstartview=FitH,colorlinks=true,linkcolor=blue,citecolor=blue,urlcolor=blue]{hyperref}
\usepackage{tikz}
\usepackage{verbatim}

\newcommand{\lettersection}[1]{\paragraph*{#1\,---\hspace{-0.29cm}}}

\usepackage[english]{babel}
\usepackage{latexsym}
\usepackage{graphics}
\usepackage{subfigure}
\usepackage{epsfig}
\usepackage{color}
\usepackage{braket}
\usepackage[T1]{fontenc}
\usepackage[latin9]{inputenc}
\usepackage{amstext}
\usepackage{esint}
\usepackage[normalem]{ulem}   

\newcommand{\beq}{\begin{equation}}
\newcommand{\eeq}{\end{equation}}

\usepackage{enumitem}
\usepackage{multirow}
\usepackage{blindtext}
\usepackage{algorithmic}
\usepackage{textcomp}
\usepackage{float}
\usepackage{array}
\newcolumntype{M}[1]{>{\centering\arraybackslash}m{#1}}
\setlist{nolistsep}
\usepackage[separate-uncertainty = true]{siunitx}
\def\BibTeX{{\rm B\kern-.05em{\sc i\kern-.025em b}\kern-.08em
    T\kern-.1667em\lower.7ex\hbox{E}\kern-.125emX}}
\PassOptionsToPackage{hyphens}{url}\usepackage{hyperref}

\usepackage{cleveref}

\begin{document}

\title{Particle-hole origin of thermal beating in dipole-compression modes of a 1D Bose gas}

\author{Caroline Mauron}
\affiliation{School of Mathematics and Physics, University of Queensland, Brisbane, Queensland 4072, Australia}

\author{Karen V. Kheruntsyan}
\email{karen.kheruntsyan@uq.edu.au}
\affiliation{School of Mathematics and Physics, University of Queensland, Brisbane, Queensland 4072, Australia}

\author{Giulia De Rosi}
\email{giulia.de.rosi@upc.edu}
\affiliation{Departament de F\'isica, Universitat Polit\`ecnica de Catalunya, Campus Nord B4-B5, 08034 Barcelona, Spain}

\date{\today}

\begin{abstract}
Using generalized hydrodynamics, we study the thermal behavior of dipole-compression collective oscillations in a harmonically trapped one-dimensional (1D) Bose gas across the crossover from weak to strong repulsive contact interactions. A key scale controlling this behavior is the temperature of the hole-induced anomaly, associated with the thermal population of hole excitations. In contrast to classical hydrodynamics, which predicts a single oscillation mode, we find a beating signal composed of two frequencies. As the temperature increases, both frequencies evolve from the low-temperature phononic hydrodynamic regime toward the collisionless limit around the anomaly temperature, without saturating at the values expected in the high-temperature collisional hydrodynamic regime. The lower frequency originates from hole excitations and is associated to low-energy oscillations, while the higher frequency emerges from particle excitations and corresponds to the dipole-compression mode. The thermal evolution of the relative excitation strengths of the two frequencies reflects the changing population imbalance between particle and hole spectral states across the anomaly. Our results reveal direct connections between excitations, thermodynamics, correlations, dynamics, and interparticle collisions, and may prove relevant to other atomic, nuclear, solid-state, electronic, and spin systems exhibiting similar anomalies or thermal second-order phase transitions. 
\end{abstract}

\maketitle

\lettersection{Introduction}

Many-body systems are ubiquitous across a broad range of disciplines, including chemistry, materials science, condensed matter, and nuclear, molecular, and atomic physics \cite{Bloch2008, Coleman2015}. By tuning the interparticle interaction strength and temperature, one can explore diverse phases of matter and regimes,  in which the system properties change at macroscopic (e.g., thermodynamics, collective dynamics) and microscopic (e.g., excitations, correlations, and collisions) levels. Remarkably, the large-scale motion of classical and quantum systems ---e.g., nuclear matter \cite{Broglia1977, Kobyakov2013}, solids \cite{Fleming1976}, magnetic materials \cite{Halperin1969}, liquid helium \cite{Landau1941}, and ultracold atomic gases \cite{Pethick2008, Griffin2009book, DeRosi2015, Pitaevskii2016}--- is often captured by classical hydrodynamics (CHD) \cite{Landau2013fluid}.

At low temperatures $T$, CHD equations describe the dynamics of a variety of systems, provided that the low-momentum (long-wavelength) quasiparticle excitations are phononic in nature, as in the one-dimensional (1D) Bose gas \cite{Lieb1963II, Menotti2002, DeRosi2017}. In a harmonic trap, phonons manifest as collective normal modes with discrete frequencies, involving slow variations of the system's macroscopic properties rather than fast microscopic particle motion. CHD is valid for any superfluid at $T\!=\!0$ K \cite{Menotti2002, DeRosi2015}, including Bose and Fermi gases and strongly interacting liquid helium \cite{Pitaevskii2016}. In 1D, any thermal phase transition, such as superfluidity and Bose-Einstein condensation (BEC), is forbidden in the thermodynamic limit \cite{Mermin1966, Hohenberg1967, Landau2013}, and CHD equations remain applicable at finite $T$ without requiring the description of coupled superfluid and normal components (Landau's two-fluid theory) \cite{Landau2013, Pitaevskii2016}.~The same occurs in an ideal Fermi gas \cite{Minguzzi2001, DeRosi2015}, where superfluidity is absent due to the lack of interactions.

At $T > T_c$ \cite{Griffin1997, DeRosi2015} (where $T_c$ is the critical temperature of the phase transition), CHD may still apply, provided that rapid and frequent interparticle collisions drive the system toward local thermal equilibrium. The high-$T$ \emph{collisional} hydrodynamic regime is expected to hold when $\omega \ll \Gamma$ \cite{Pethick2008, Pitaevskii2016}, where $\omega$ is the collective mode frequency and $\Gamma$ the collision rate. As temperature increases further, the density decay induced by the harmonic trap makes collisions rare and any gas inevitably ends up in the \emph{collisionless} (non-hydrodynamic) regime (see Appendix A), captured by the classical ideal gas model.

In atomic, solid-state, electronic, and spin systems, the specific heat may show a peak ---referred to as an \textit{anomaly}--- located at temperature $T_A$ \cite{DeRosi2022}.~The specific heat is defined as the variation of the internal energy with temperature.~Consequently, the anomaly signals energy gaps or states thermally populated in the excitation spectrum \cite{Tari2003, He2009, DeRosi2022}.~Anomalies may appear at the onset of a thermal second-order phase transition (e.g., superfluidity or BEC) at $T_c \!= \!T_A$ \cite{Raju1992, Ku2012, Landau2013}.~In 1D gases, where phase transitions are forbidden, the peak in the specific heat arises whenever phonons are present \cite{DeRosi2022}.~The \textit{Schottky anomaly} occurs in systems with a discrete energy spectrum, such as a two-level model, when the thermal energy becomes comparable to the gap between the lowest levels. The \textit{hole-induced anomaly} was identified in a uniform 1D Bose gas at arbitrary interaction strength and originates from the thermal occupation of a region of empty states --simulating an energy gap-- lying below the branch of hole excitations in the spectrum \cite{DeRosi2022}. At $T \approx T_A$, anomalies also manifest in other thermodynamic quantities \cite{DeRosi2022, DeRosi2024}, signaling the breakdown of the low-temperature quasiparticle description of excitations \cite{Yan2020, DeRosi2022}. At $T\! \gtrsim \!T_A$, thermal fluctuations prevail over quantum effects, modifying the correlation properties of the system \cite{DeRosi2022,DeRosi2024}.  

This Letter sheds light on the impact of the hole-induced anomaly on collisional effects and collective oscillations in harmonically trapped configurations, thereby providing a universal framework that connects microscopic and macroscopic phenomena in 1D Bose gases.~In doing so, we investigate dipole-compressional (DC) oscillations \cite{Tey2013,Hu2014,DeRosi2015,DeRosi2016} --the lowest compressional modes with the same parity as the center-of-mass (dipole) mode-- across the anomaly temperature, explaining their behavior in terms of the thermal population of the underlying excitations in the spectrum.~To this end, DC modes provide excellent probes, since their frequencies at high temperatures differ between the collisional hydrodynamic \cite{Hu2014} and collisionless regimes \cite{DeRosi2015, DeRosi2016}, as shown in Table~\ref{Tab:DC freq}.~This behavior contrasts with that of the lowest breathing mode---extensively studied experimentally and theoretically
\cite{Sinatra2001, Menotti2002, Moritz2003, Haller2009, Schmitz2013, Fang2014, Hu2014, Chen2015, DeRosi2015, Gudyma2015, Choi2015, DeRosi2016, Bouchoule2016, Bayocboc2023}---which has the same frequency in both regimes.~The 1D Bose gas can be realized in ultracold atom chips, optical lattices and tube traps \cite{Tolra2004, Kinoshita2004, Bouchoule2011, SalcesCarcoba2018}, making it an ideal platform to study DC modes, which have been measured in Fermi gases \cite{Tey2013}.

To  investigate the DC oscillations, we employ the theory of generalized hydrodynamics (GHD) \cite{Castro-Alvaredo2016,Bertini2016}, which extends beyond CHD \cite{Doyon2017_PRL,Doyon2017,Schemmer2019,Doyon2020,Bastianello2021,Alba2021,Malvania2021, Kerr2023, Watson2023} as it is capable of capturing also the collisionless regime.~Our GHD simulations thus serve as a numerical experiment, and provide evidence for the absence of a high-$T$ collisional hydrodynamic regime which we explain with the breakdown of the quasiparticle picture induced by the anomaly.~Our results reveal a thermal crossover from \emph{phononic} hydrodynamic to \textit{collisionless} behavior around the anomaly temperature $T_A$ \cite{DeRosi2022}. We therefore propose that $T < T_A$ serves as a sufficient and universal criterion for the applicability of CHD, restricting it only to the phononic hydrodynamic regime.

More generally, our work suggests that anomalies can serve as probes to tackle the challenging many-body problem, enabling the prediction of collective dynamics from excitations across a wide variety of finite-temperature systems, from atomic to nuclear matter.

\begin{table}[tbp]
\centering
\caption{Hydrodynamic vs. collisionless frequencies of dipole compression modes in a 1D Bose gas for weak and strong interactions \cite{DeRosi2015, DeRosi2016}. $\omega_x$ is the axial harmonic trap frequency.}
\setlength\extrarowheight{3pt}
\begin{tabular}{c|cc|c}
\hline
 \hline
& \multicolumn{2}{c|}{Hydrodynamic} & \multirow{2}{*}{Collisionless} \\
& $T=0$ K \,\,\,\,\,\,\,\,\,\,\,\,& high $T$ &   \\
\hline
Weak interactions \,\,& $\sqrt{6} \omega_x$ \,\,\,\,\,\,\,\,\,\,\,\,& $\sqrt{7} \omega_x$ & $3 \omega_x, 1 \omega_x$ \\ 
\hline
Strong interactions & $3 \omega_x$ \,\,\,\,\,\,\,\,\,\,\,\,& $\sqrt{7} \omega_x$ & $3 \omega_x, 1 \omega_x$ \\
 \hline
 \hline
\end{tabular}
\label{Tab:DC freq}
\end{table}

\lettersection{Model}

The Hamiltonian for the harmonically trapped 1D Bose gas is given by \cite{Petrov2000,Kheruntsyan2005} 
\begin{multline}
\label{Eq:H}
H\! =\! -\frac{\hbar^2}{2m}\! \sum_{i = 1}^N \frac{\partial^2}{\partial x_i^2} + g_{\rm 1D}\! \sum_{i > j}^N\!\delta(x_i \!-\! x_j) + \!\sum_{i = 1}^N\!V(x_i),
\end{multline} 
where $m$ is the particle mass, $N$ is the total number of bosons, and $V(x) \!=\! m\omega_x^2 x^2/2$ is the axial harmonic trap with frequency $\omega_x$. The interaction strength is $g_{\rm 1D}
\! \simeq \!2 \hbar \omega_\perp a$ \cite{Olshanii1998} for repulsive ($a\!>\!0$) contact interactions, with $\omega_{\perp}$ the transverse harmonic confinement frequency and $a$ the three-dimensional (3D) $s$-wave scattering length.

In the absence of the trap, $V(x)\!=\!0$, the density $n\!=\!N/L$ (where $L$ is the length of the system) is constant in $x$ and the Hamiltonian reduces to the integrable Lieb-Liniger model \cite{Lieb1963, Lieb1963II}. This framework is exactly solvable via the Bethe ansatz \cite{Bethe1931, Takahashi1999} for its $T\!=\!0$ ground state and the spectrum of excitations, as well as for  finite-$T$ thermodynamic properties using Yang-Yang's thermal Bethe ansatz \cite{Yang1969, Yang1970,Kheruntsyan2003}. When $V(x)\!\ne\!0$, the system becomes non-integrable, and the spatial density profile $n(x)$ determines $N\!=\! \int dx \, n(x)$. Under weak confinement, which slightly breaks integrability, the gas is approximately uniform locally, allowing the use of the local density approximation \cite{Kheruntsyan2005}, where $n(x)$ follows from the local chemical potential $\mu[n(x)]\!=\! \mu_0 - V(x)$ with $\mu_0 \!=\! \mu(x\!=\!0)$.

\begin{figure}[t]
\includegraphics[width=0.98\columnwidth]{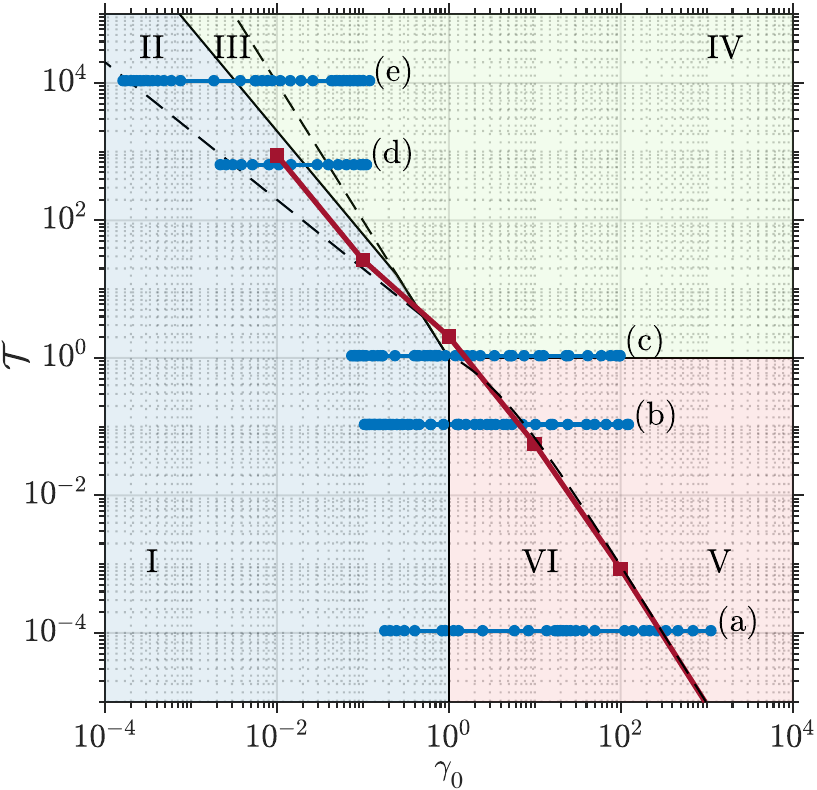}
\caption{Diagram of regimes of the harmonically trapped 1D Bose gas in terms of the interaction strength $\gamma_0$ and temperature $\mathcal{T}$ \eqref{Eq:gamma and T parameters}.~Regimes are \cite{Kerr2024}:~I.~Bogoliubov (quantum), $\mathcal{T}/2 \ll \gamma_0^{-1}, \gamma_0 \ll 1$; II.~Bogoliubov (thermal), $2 \gamma_0^{-1} \ll \mathcal{T} \ll 2\gamma_0^{-3/2}$; III.~Nearly ideal Bose gas (degenerate), $2\gamma_0^{-3/2}\ll \mathcal{T}\ll \gamma_0^{-2}$; IV.~Nearly ideal Bose gas (non-degenerate), $\mathcal{T} \gg \text{max}\{1,\gamma_0^{-2}\}$; V.~Strong interactions (high $T$), $\pi^2/\left(  \gamma_0 + 2\right)^2 \ll \mathcal{T} \ll 1$, $\gamma_0 \gg 1$; VI.~Strong interactions (low $T$), $\mathcal{T} \ll \pi^2/\left(  \gamma_0 + 2\right)^2$, $\gamma_0 \gg 1$. Red thick solid line denotes the anomaly temperature $\mathcal{T}_A$ vs. $\gamma_0$, where the red square markers represent the values of $\mathcal{T}_A$ estimated from the position of the peak in the specific heat \cite{DeRosi2022}.~Datasets (a)-(e) (blue points), are used for the study of dipole compression modes in regimes I-VI, and are calculated at fixed $\mathcal{T}$. 
}
\label{Fig:Regimes_Diagram}
\end{figure}

The system \eqref{Eq:H} is conveniently characterized by the dimensionless interaction strength and temperature \cite{Kheruntsyan2005}:
\begin{equation}
\label{Eq:gamma and T parameters}
\gamma_0 = \frac{m g_{\rm 1D}}{\hbar^2 n_0}, \quad  \mathcal{T}=\frac{T/T_d}{\gamma_0^2} = \frac{2 \hbar^2 k_B T}{m g_{\rm 1D}^2},
\end{equation}
where $T_d \!=\! \hbar^2 n_0^2/(2m k_B)$ is the quantum degeneracy temperature at the central density $n_0\! =\! n(x \!= \!0)$.

Figure~\ref{Fig:Regimes_Diagram} shows the diagram of different regimes of the 1D Bose gas \cite{Petrov2000, Kheruntsyan2003,Kheruntsyan2005, DeRosi2019, DeRosi2022, DeRosi2023, Kerr2024} in the $\gamma_0$--$\mathcal{T}$ parameter space. The points on the horizontal lines (a)--(e) depict the datasets of this work. Progressing to the right along a given line corresponds to increasing values of the interaction strength at the trap center $\gamma_0$,  associated with decreasing densities $n_0$. Datasets (a)--(e) span all relevant physical regimes, including the weakly interacting quasicondensate regions I and II, described by Bogoliubov theory \cite{Castin2004}; the degenerate (III) and non-degenerate (IV) regimes, both approaching the ideal ($\gamma_0\!\to \!0$) Bose gas; and the strongly interacting high-$T$ (V) and low-$T$ (VI) behaviors, which for $\gamma_0 \!\to \!\infty$ recover the Tonks-Girardeau limit, where the thermodynamics is the same as that of an ideal Fermi gas \cite{Girardeau1960}. Figure~\ref{Fig:Regimes_Diagram} also reports the dimensionless hole-induced anomaly temperature $\mathcal{T}_A$ as a function of $\gamma_0$ \cite{DeRosi2022}, which intersects datasets (a)--(d). The values of $\mathcal{T}_A$ (red squares) are obtained from the peak of the specific heat in the uniform system with the same $\gamma_0$ as in the trap. For dataset (e), which lies in the very weakly interacting regime ($\gamma_0 < 10^{-1}$), the peak is barely discernible, so no $\mathcal{T}_A$ is shown.

For fixed $\gamma_0$ and $\mathcal{T}$, a third parameter is needed to fully describe a confined gas---typically the trap frequency $\omega_x$, which sets the atom number $N$, or vice versa. In Appendix B, we report $N$ as a function of $\gamma_0$ for datasets (a)--(e) at fixed $\omega_x$, along with the parameters of our calculations of the DC frequencies. We also discuss the experimental conditions required to access regimes I-VI.

\lettersection{Dipole Compression Modes}

\begin{figure*}[tbp]
\centerline{\includegraphics[width=0.99\textwidth]{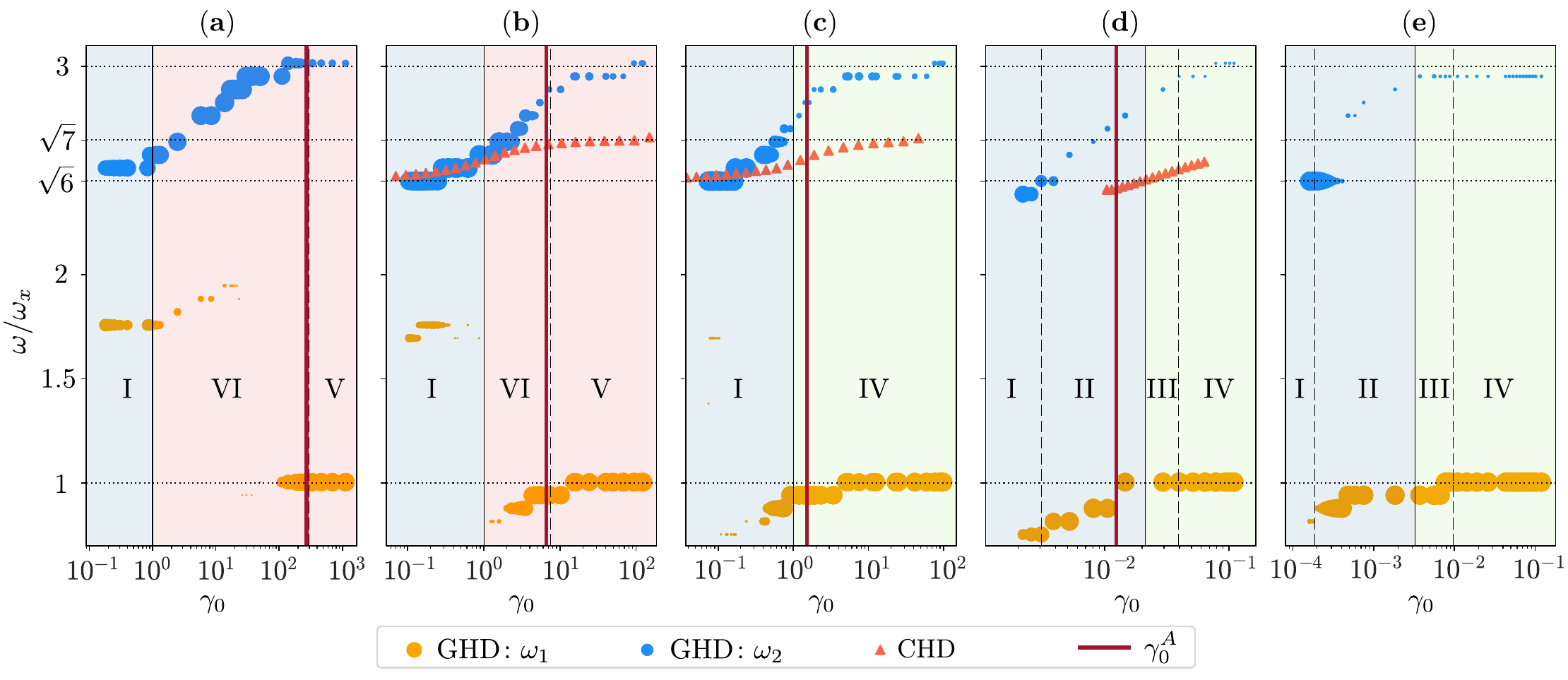}}
\caption{Frequencies $\omega/\omega_x$ of the dipole compression modes vs. the interaction strength $\gamma_0$, Eq.~\eqref{Eq:gamma and T parameters}, for datasets (a)-(e), shown in the panels, which span the regimes I-VI of the 1D Bose gas of Fig.~\ref{Fig:Regimes_Diagram}.~Additional parameter values of our simulations are listed in Appendix B, Table \ref{Tab:Parameters}. The two dominant frequency components, $\omega_1$ (lower) and $\omega_2$ (higher), are extracted from generalized hydrodynamics (GHD) calculations of the Fourier transform of the skewness (Appendix D). The sizes of the respective yellow and blue markers are proportional to the excitation strengths $K_1$ and $K_2$.~Red triangles indicate the predictions from Ref.~\cite{Hu2014} obtained using a classical hydrodynamic (CHD) variational ansatz.~Dotted horizontal lines at $\omega/\omega_x=\{1,\sqrt{6},\sqrt{7},3\}$ mark the analytic limits of Table \ref{Tab:DC freq}.~Vertical solid red lines denote the interaction strength at the hole-induced anomaly temperature $\gamma_0^A \!\equiv \!\gamma_0\!\left(\mathcal{T}\!=\! \mathcal{T}_A\right)$, whose values are: (a) $\gamma_0^A\!=\!2.86 \times 10^2$; (b) $6.57$; (c) $1.51$; (d) $1.23 \times 10^{-2}$.}
\label{Fig:DC frequencies}
\end{figure*}

We initialize the system, Eq.~\eqref{Eq:H}, in a finite-$T$ equilibrium state using the exact thermal Bethe ansatz within the local density approximation \cite{Kheruntsyan2005}.~To excite DC oscillations, we apply at time $t = 0$ a quench to the harmonic trap potential $V(x)\rightarrow V(x)-\lambda \left(x^3/3 - x \langle x^2 \rangle\right)$ \cite{DeRosi2016}, where $\lambda$ is a small perturbation strength and $\langle x^2 \rangle$ is the average evaluated with the initial (pre-quench) equilibrium density profile $n(x)$ \cite{DeRosi2017thesis} (Appendix C).

To characterize the ensuing DC modes, we compute the time evolution of the perturbed density $n(x,t)$ using the GHD equations \cite{Castro-Alvaredo2016,Bertini2016}, which are implemented with the \textit{iFluid} package \cite{Moller2020}, as in Ref. \cite{Watson2023}.~GHD \footnote{
See Supplemental Material at [...] for a summary of the formalism of the generalized hydrodynamics. The Supplemental Material includes Refs.~\cite{Thacker1981, Yang1969, Sutherland2004, Rigol2007, Rigol2008, cazalilla2010focus, Polkovnikov2011, Eisert2015, Gogolin2016, Castro-Alvaredo2016, Bertini2016, Doyon2017, Doyon2017_PRL, Schemmer2019, Doyon2020, Moller2020, Bastianello2021, Alba2021, Malvania2021, Kerr2023, Watson2023}} describes the collective dynamics of integrable systems out of thermal equilibrium.~Even when integrability is weakly broken, GHD remains applicable within the local density approximation \cite{Kheruntsyan2005, Doyon2017, Bastianello2021, Watson2023}.~GHD provides accurate predictions at length scales much larger than the average interparticle distance, where the gas can be modelled as a continuous fluid in local thermal equilibrium and has been benchmarked and validated in numerous theoretical and experimental studies \cite{Doyon2017_PRL,Doyon2017,Schemmer2019,Doyon2020,Bastianello2021,Alba2021,Kerr2023}.

Once $n(x,t)$ is computed over a sufficiently long time period, we extract the frequencies of the DC oscillations from the Fourier spectrum of the skewness of $n(x,t)$ (see Appendix D).~The resulting spectra typically exhibit two dominant peaks, indicating a beating of two DC frequencies; the positions of the peaks correspond to the said frequencies, $\omega_1$ and $\omega_2 > \omega_1$. We also characterize their relative excitation strengths $K_{1,2}$ via the peak heights $h_{1,2}$, using $K_1 = h_1^2/(h_1^2 + h_2^2)$ and $K_2=1-K_1$ \cite{Bayocboc2023}.

Figure \ref{Fig:DC frequencies} summarizes our GHD results for the DC frequencies $\omega_{1,2}$ and their $K_{1,2}$ as functions of the interaction strength $\gamma_0$, Eq.~\eqref{Eq:gamma and T parameters}.~The different panels correspond to datasets (a)--(e) (see Table \ref{Tab:Parameters}, Appendix B).~For comparison, we also present the predictions of the CHD variational ansatz \cite{Hu2014} as red triangles.~For all (a)--(e), increasing $\gamma_0$ is equivalent to raising the temperature, as the system is driven into the two high-$T$ regimes IV and V, see Fig.~\ref{Fig:Regimes_Diagram}.~In panels (a)--(d), vertical red lines mark the values of $\gamma_0$ at the anomaly temperature, $\gamma_0^A \!\equiv \!\gamma_0(\mathcal{T} \!=\! \mathcal{T}_A)$, where $\mathcal{T}_A$ intersects (a)--(d) in Fig.~\ref{Fig:Regimes_Diagram}.

For $\gamma_0\!>\!\gamma_0^A$, equivalent to temperatures $T\!> \!T_A$, the GHD simulations confirm the existence of the collisionless regime, identified by the beating of two frequency components, $\omega_1\!\simeq \!1\omega_x$ and $\omega_2\!\simeq \!3 \omega_x$ \cite{DeRosi2016}. However, unlike Ref.~\cite{DeRosi2016}, which predicted equal excitation strengths for these frequencies (Table \ref{Tab:DC freq}), we find that the low-energy component $\omega_1$ dominates $\omega_2$ as $K_1 \!>\! K_2$.~This discrepancy is explained by the hole-induced anomaly, arising from the thermal occupation of states located below the lowest hole branch in the spectrum \cite{DeRosi2022}.~The excitation of these low-energy hole states is confirmed by the behavior of the dynamic structure factor (DSF) at $T \!> \!T_A$ \cite{DeRosi2022}.

For $\gamma_0 \!<\! \gamma_0^A$ ($T\!<\! T_A$), we instead observe that $\omega_2$ dominates over $\omega_1$, as $K_2 \!>\! K_1$. This is again consistent with the behavior of the DSF at $T \!<\! T_A$, which reveals the excitation of high-energy particle states located above the hole spectral branch \cite{DeRosi2022}.~Our GHD results for the $\omega_2$ component agree with the predictions of the CHD variational ansatz, and both recover the analytical limit $\omega_2 \!=\! \sqrt{6}\omega_x$ of the Bogoliubov regime I at $\gamma_0 \!\ll \!\gamma_0^A$ ($T\!\ll \! T_A$), thereby confirming the existence of the phononic hydrodynamics at low $T$ (see Table \ref{Tab:DC freq}).

We therefore conclude that the anomaly temperature $T_A$ defines a new energy scale governing the hydrodynamic-to-collisionless crossover.~Since $T_A$ depends on $\gamma_0$ \cite{DeRosi2022}, the single universal condition $T < T_A \left(\gamma_0\right)$ is sufficient to determine the phononic hydrodynamic regime for any interaction strength.~This contrasts with the corresponding two independent conditions in $\gamma_0$ and $T$ of Ref. \cite{Hu2014}.~Physically, the relevance of the anomaly mechanism for the validity of the hydrodynamic regime is confirmed by the breakdown of the effective low-temperature quasiparticle picture at $T \approx T_A$, as evidenced by the DSF \cite{DeRosi2022}.~Notably, a similar trend is observed in systems of higher spatial dimensionality undergoing superfluid or BEC phase transitions:~when crossing the critical temperature $T_c$, the DSF reveals a transition from well-defined collective quasiparticle excitations at $T < T_c$ to a broad thermal response for $T > T_c$.

Next, the findings of the CHD variational ansatz from Ref.~\cite{Hu2014} disagree with our GHD results.~Qualitatively, such a CHD method predicts a single frequency across all regimes, whereas GHD reveals a beating pattern involving two distinct frequencies,  $\omega_1$ and $\omega_2$ --corresponding to hole and particle spectral states, respectively-- whose excitation strengths $K_{1,2}$ vary with $\gamma_0$. Quantitatively, at $\gamma_0 \!\gg \!\gamma_0^A$ ($T \gg T_A$), the oscillation frequency of the CHD variational ansatz asymptotically approaches the value $\sqrt{7}\omega_x$, corresponding to the prediction for the high-temperature collisional hydrodynamic regime in Table \ref{Tab:DC freq}. However, our GHD results show that $\omega_2$ transitions from $\sqrt{6}\omega_x$ through $\sqrt{7}\omega_x$ at intermediate $\gamma_0$---without saturating there---and continues to increase until it reaches the collisionless limit $3\omega_x$ at very large $\gamma_0$. These observations confirm: (i) the absence of the high-$T$ collisional hydrodynamic regime, in contrast with previous theoretical predictions \cite{Hu2014, DeRosi2015, DeRosi2016}; and (ii) the failure of classical hydrodynamic approaches---including CHD variational ansatz---to capture the collisionless behavior, which instead requires other frameworks like the GHD.

Finally, in panels (a)--(c), for $\gamma_0 \!\ll \!\gamma_0^A$ ($T\!\ll \! T_A$), the GHD frequency $\omega_1$ recovers the phononic hydrodynamic limits for the lowest breathing (LB) mode, with values $\omega_1\!=\!\sqrt{3} \omega_x$ and $\omega_1\!=\!2 \omega_x$, corresponding to the low-temperature regimes I and VI, respectively \cite{Fang2014,Hu2014, DeRosi2015,DeRosi2016,Bouchoule2016}.~This result is remarkable, as
our GHD simulations were performed without applying the specific LB-mode perturbation to the trap, $V(x)\rightarrow V(x)-\lambda \left(x^2 - \langle x^2 \rangle\right)$ \cite{DeRosi2016}. This suggests that, in the present DC oscillations, the LB mode emerges naturally at very low $T$, with $\omega_1$ being coupled to $\omega_2$.~Moreover,  in all panels at $\gamma_0 \!\sim\! \gamma_0^A$, we even see $\omega_1 \!<\! \omega_x$---a mode at low excitation energy, which also appears in the DSF at $T \sim T_A$ \cite{DeRosi2022}.~As $\gamma_0$ (or the temperature) increases, the LB oscillation gradually weakens and eventually vanishes in favour of the mode with $\omega_1 < \omega_x$.

\lettersection{Conclusions}

We presented a systematic numerical study of the thermal behavior of dipole compression mode frequencies in a harmonically trapped 1D Bose gas with arbitrary interaction strength. Our results: (i) reveal a two-frequency beating between particle- and hole-like modes, whose excitation strengths track the thermal occupation of spectral states induced by the hole anomaly \cite{DeRosi2022}; (ii) show a crossover from low-$T$ phononic hydrodynamic behavior to a high-$T$ collisionless regime around the anomaly temperature $T \approx T_A$; (iii) indicate that the lower frequency corresponds to the unexpected excitation of the lowest breathing mode at low temperatures.

An anomaly-induced suppression of collisional hydrodynamics is expected in other 1D systems, such as the super Tonks-Girardeau gas \cite{Astrakharchik2005, Haller2009}, dipolar \cite{Arkhipov2005, Girardeau2012} and Rydberg \cite{Osychenko2011} ensembles, ${^3}$He gas \cite{Astrakharchik2014}, liquid ${^4}$He \cite{Bertaina2016, DelMaestro2022}, liquids in mixtures \cite{Cheiney2018, DeRosi2021}, and spin chains \cite{Bertini2016}.~It is intriguing to explore systems where the anomaly manifests as a superfluid or BEC phase transition, such as in Fermi \cite{Ye2025} and Bose \cite{Hiyane2024} atomic gases, and in nuclear matter in neutron stars \cite{Haensel1976, DeRosi2013thesis, Benhar2017}.


\vspace{0.1cm}

\begin{acknowledgments} 

\lettersection{Acknowledgments}

The authors thank H.~Hu and X.-J.~Liu for kindly sharing the CHD variational ansatz data from Ref.~\cite{Hu2014} shown in Fig.~\ref{Fig:DC frequencies} for comparison purposes. C.M. is grateful for the computational support of I. Mortimer and the Core Computing Facility ``Getafix'' of the School of Mathematics and Physics at the University of Queensland. K.V.K. acknowledges support by the Australian Research Council Discovery Project Grant No.~DP240101033. G.D.R. received funding from the Universitat Polit\`ecnica de Catalunya (ALECTORS2024, R-02673) and the grant IJC2020-043542-I funded by MCIN/AEI/10.13039/501100011033 and by "European Union NextGenerationEU/PRTR".~G.D.R. also acknowledges support by the Spanish Ministerio de Ciencia, Innovaci\'on y Universidades (grant PID2023-147469NB-C21, financed by MICIU/AEI/10.13039/501100011033 and FEDER-EU).
\end{acknowledgments}


\vspace{0.1cm}

\lettersection{Author Contributions}
G. D. R. devised the initial concepts and the theoretical framework. 
C.M. performed the GHD numerical simulations under the supervision of K.V.K.
All authors contributed to the analysis, interpretation and visualization of the results.
G.D.R. and K.V.K. jointly wrote the manuscript. 

\vspace{0.1cm}

\lettersection{Appendix A: High-$T$ collisionless regime}

At sufficiently high temperatures $T$, any harmonically trapped gas approaches the ideal classical limit, as the confinement causes a rapid decay in the density profile, which is described by the Maxwell-Boltzmann Gaussian. 

For the sake of generality, consider a classical gas of $N$ particles in a three-dimensional (3D) isotropic harmonic trap with frequency $\omega_{\rm ho}$. The corresponding Maxwell-Boltzmann density profile is given by $n_{\mathrm{3D}}\left(r\right) = \frac{N}{\pi^{3/2} R_T^3} e^{-r^2/R_T^2}$, where $r = \sqrt{x^2 + y^2 + z^2}$ is the scalar radial coordinate and $R_T = \sqrt{2 k_B T/\left(m \omega_{\rm ho}^2\right)}$ is the thermal radius \cite{Pitaevskii2016}. The collision rate in such a gas is given by $\Gamma = n_{\mathrm{3D}}\left(r\right) \sigma v_T$ and depends on the $s$-wave scattering cross-section $\sigma = 8 \pi a^2$ (where $a$ is the 3D $s$-wave scattering length) and the thermal velocity $v_T  \sim \sqrt{T}$.~At high $T$, collisions become rare as the collision rate $\Gamma$ decreases with temperature according to $\Gamma \sim \sigma v_T/R_T^3 \sim \sigma/T$ \cite{DeRosi2017thesis}. Consequently, a harmonically trapped (i.e., nonuniform) gas necessarily enters the collisionless regime,  as $\Gamma\to 0$ with $T\to\infty$.  In contrast to this,  in a uniform system, where the density of the gas $n_{\mathrm{3D}}$ remains constant in $r$ at any $T$, collisions become more frequent as $\Gamma$ increases with temperature through its dependence on $v_T$, which is unchanged.

\lettersection{Appendix B: Parameter values}

For most of the datasets (a)--(e) in Fig.~\ref{Fig:Regimes_Diagram}, our simulations are performed using realistic experimental parameters \cite{Moritz2003, Tolra2004, Kinoshita2004,Kinoshita2005,Hofferberth2007,Haller2009,Kruger2010,Armijo2010, Bouchoule2011, Gring2012,Fang2014,Schemmer2019,Shah2023}, with temperatures in the range $T \simeq 60 \div 1000$ nK, total number of atoms $N \gtrsim 80$, and axial and transverse (radial) trap frequencies of $\omega_x/2 \pi \simeq1\div 10$ Hz and $\omega_\perp/2\pi \simeq 400 \div 2\times 10^4$ Hz, respectively. To access some of the more extreme physical regimes, however, we had to push the values of these parameters beyond what is currently achievable in experiments. We recall that entering the 1D regime requires $\omega_x \ll \omega_\perp$, along with that both the central chemical potential $\mu_0$ and the thermal energy $k_B T$ satisfy $\{\mu_0, k_BT\}\ll\hbar\omega_{\perp}$ \cite{Kruger2010,Shah2023}.

We consider $^{87} \rm{Rb}$ atoms (with mass $m\simeq 1.443 \times 10^{-25}$ kg) for definiteness, for which the 3D $s$-wave scattering length is $a \simeq 5.3$ nm. The value of $a$ can be tuned via magnetic Fano-Feshbach resonances \cite{Chin2010, Meinert2015}, which in turn changes the 1D interaction strength $g_{\mathrm{1D}} \simeq 2 \hbar \omega_\perp a$ away from confinement-induced resonances \cite{Olshanii1998,Haller2009}. Alternatively, $g_{\mathrm{1D}}$ can be tuned by varying $\omega_\perp$ \cite{Olshanii1998, Kinoshita2004, Kinoshita2005}. We note, however, that close to a confinement-induced resonance, where $a \to a_\perp/C$, one must use the exact expression $g_{\rm 1D}=2\hbar^2 a/[ma_{\perp}^2(1-C a/a_\perp)]$, where $a_\perp=\sqrt{\hbar/\left(m\omega_\perp\right)}$ is the transverse harmonic oscillator length and $C \simeq 1.0326$ \cite{Olshanii1998, Pitaevskii2016}.

Temperature can be measured both below and above the hole-induced anomaly threshold $T_A$ in single 1D tubes,  using techniques such as time-of-flight,  Bose gas focusing, ideal Bose gas fits to the wings of the \emph{in-situ} density profiles, or measurements of \emph{in-situ} density fluctuations \cite{vanAmerongen2008,Armijo2010,Jacquim2011,SalcesCarcoba2018}.  More recently, advances in thermometry have been achieved using neural networks, trained to extract temperature from a single absorption image taken after time-of-flight expansion \cite{Moller2021}.

\begin{table}[tbp]
\centering
\caption{Values of the dimensionless temperature $\mathcal{T}$, Eq.~\eqref{Eq:gamma and T parameters}, the dimensionless 1D scattering length $\tilde{a}_{\mathrm{1D}}$, Eq.~\eqref{Eq:a_1D}, the 3D $s$-wave scattering length $a$ (in meters m), and the dimensionless perturbation strength $\tilde{\lambda}\ll 1$, Eq.~\eqref{Eq:lambda}, for each dataset (a)-(e), corresponding to blue points in Figs.~\ref{Fig:Regimes_Diagram} and \ref{Fig:N_gamma0}, and to panels (a)-(e) in Fig.~\ref{Fig:DC frequencies}. All parameter values are chosen such that they satisfy the 1D condition $\{\mu_0, k_B T, \hbar \omega_x\} \ll \hbar \omega_\perp$.
}
\setlength\extrarowheight{3pt}
\begin{tabular}{c|c|c|c|c}
\hline
 \hline
Dataset & $\mathcal{T}$ & $\tilde{a}_{\mathrm{1D}}$ & $a$ [m] & $\tilde{\lambda}$\\
\hline
(a) & $1.1 \times 10^{-4}$ & $10^{-3}$ & $2.65 \times 10^{-8}$ & $3 \times 10^{-5}$ \\
\hline
(b) &  $0.11$ & $1.43 \times 10^{-2}$ & $2.65 \times 10^{-8}$ & $5 \times 10^{-4}$  \\
\hline
(c) & $1.08$ & $3.22 \times 10^{-2}$ & $2.65 \times 10^{-8}$ & $10^{-3}$  \\
\hline
(d) & $6.5 \times 10^2$ & $3.23$ & $5.3 \times 10^{-9}$ & $2 \times 10^{-4}$  \\
\hline
(e) & $1.1 \times 10^4$ & $3.23$ & $5.3 \times 10^{-11}$ & $3.5 \times 10^{-3}$  \\
\hline
 \hline
\end{tabular}
\label{Tab:Parameters}
\end{table}

The parameters employed in our simulations of DC oscillations in Fig. \ref{Fig:DC frequencies} for the datasets (a)-(e) are shown in Table~\ref{Tab:Parameters}. Here, the  temperature $\mathcal{T}$ is defined in Eq.~\eqref{Eq:gamma and T parameters}. The dimensionless 1D scattering length is given by:
\begin{equation}
\label{Eq:a_1D}
\tilde{a}_{\mathrm{1D}} = - a_{\mathrm{1D}}/a_x,
\end{equation}
where $a_{\rm 1D} = - 2 \hbar^2/\left(m g_{\rm 1D}\right) < 0$ is the 1D scattering length \cite{Olshanii1998} and $a_x = \sqrt{\hbar/\left(m \omega_x\right)}$ is the axial harmonic oscillator length. The dimensionless perturbation strength is defined as:
\begin{equation}
\label{Eq:lambda}
\tilde{\lambda} = \lambda a_x^3/(\hbar \omega_x).
\end{equation}

As mentioned in the main text, for fixed values of the interaction strength $\gamma_0$ and temperature $\mathcal{T}$, Eq.~\eqref{Eq:gamma and T parameters}, a third parameter must be specified to fully and uniquely characterize a harmonically trapped 1D Bose gas. This can either be the frequency of the axial confinement $\omega_x$, which fixes the total number of atoms $N$ at a given central density $n_0$ (or  $\gamma_0$), or, alternatively, $N$ itself,  in which case we treat $\omega_x$ as an adjustable input parameter.

\begin{figure}[tbp]
\includegraphics[width=0.98\columnwidth]{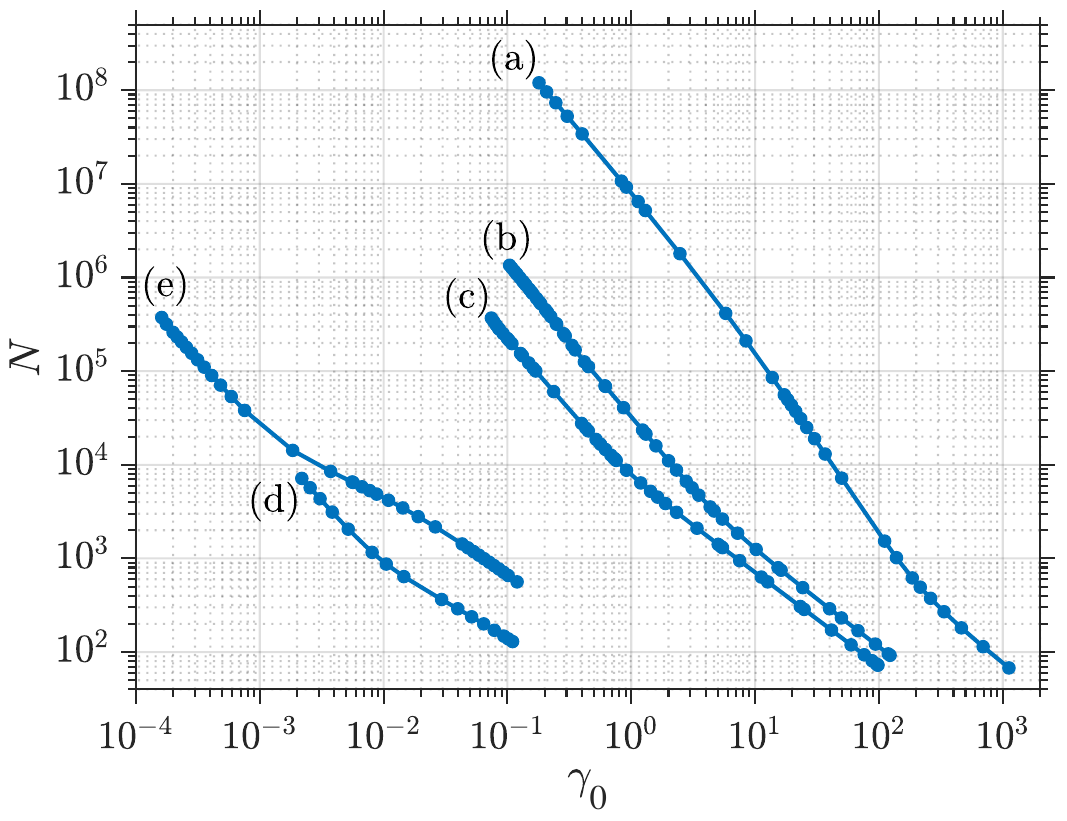}
\caption{Total atom number $N$ of a harmonically confined 1D Bose gas vs. the interaction strength $\gamma_0$, Eq.~\eqref{Eq:gamma and T parameters}, defined at the trap centre.~Each dataset (a)-(e) (see also Fig.~\ref{Fig:Regimes_Diagram}) is shown for a fixed value of both temperature $\mathcal{T}$, Eq.~\eqref{Eq:gamma and T parameters}, and the 1D scattering length $\tilde{a}_{\mathrm{1D}}$, Eq.~\eqref{Eq:a_1D}, as in Table \ref{Tab:Parameters}. 
}
\label{Fig:N_gamma0}
\end{figure}

In Fig.~\ref{Fig:N_gamma0} we report the values of $N$ as a function of $\gamma_0$ that we used for each of the datasets (a)-(e) at their respective fixed $\mathcal{T}$ and $\tilde{a}_{\mathrm{1D}}$ as in Table \ref{Tab:Parameters}. By keeping $\mathcal{T}$ constant, we fix the value of $g_{\rm 1D}$ which---together with the condition of constant $\tilde{a}_{\mathrm{1D}}$---allows us to determine $\omega_x$.   
For each dataset (a)--(e) in Table \ref{Tab:Parameters}, we also keep fixed both $a$, which defines $\omega_\perp$ via $g_{\rm 1D}= 2\hbar\omega_\perp a$, and the perturbation strength $\tilde{\lambda}$, which is required to excite the dipole-compressional modes.

\lettersection{Appendix C: Excitation of DC modes}
Collective oscillations are  generated by a perturbation $H_{\rm pert} = -\lambda F(x) \theta(t)$ to the Hamiltonian \eqref{Eq:H} with a small strength  parameter $\lambda$.  Here,  $F(x) = \sum_{i = 1}^N f\left(x_i\right)$ is the static excitation operator and $\theta(t)$ is the Heaviside function of time $t$.~$H_{\rm pert}$ is implemented as a sudden quench of the external trapping potential $V(x)\rightarrow V(x)-\lambda f(x)$ at $t = 0$ leading to a density perturbation $n(x, t)$.

To excite the dipole compression modes, we choose $f(x)$ as $f_{\rm DC}(x) = x^3/3 - x \langle x^2 \rangle$,  where $\langle x^2 \rangle = \int dx\, x^2\,  n(x)/N$ is the average evaluated using the equilibrium density $n(x)$ prior to the quench \cite{DeRosi2017thesis}. The subtraction of the second term in $f_{\rm DC}(x)$ prevents the excitation of the center-of-mass (dipole) mode \cite{DeRosi2016} whose frequency is $\omega_x$.

\lettersection{Appendix D: Skewness and its Fourier Transform}

To characterize the DC oscillations in the perturbed density $n(x,t)$, we employ the skewness as our probe, which describes the asymmetry of the atomic distribution in the gas cloud following the quench. It is defined as the standardized (i.e., normalized by the standard deviation) third moment of the distribution of atom positions $x$: 
\begin{equation}
\label{Eq:Skew}
\mathrm{Skew}(t) = \frac{\langle  \left(x - \langle x \rangle_t    \right)^3\rangle_t}{\sigma_x^3\left(t\right)},
\end{equation}
where $\sigma_x(t) = \sqrt{\langle x^2 \rangle_t - \langle x \rangle_t^2}$ is the standard deviation, and the averages $\langle \cdot \rangle_t$ are evaluated with the instantaneous  density $n(x,t)$, which is computed using the generalized hydrodynamics method.  We note that the third moment of the atom number distribution in a 1D Bose gas was measured in Ref.~\cite{Armijo2010},  suggesting that skewness could likewise be accessed in state-of-the-art experiments probing the same system.

We analyse $\mathrm{Skew}(t)$ in frequency space by performing its discrete Fourier transform (FT), which is real for the norm $\Vert\cdot \Vert$, and is given by:
\begin{equation}
\label{Eq:FT Skew}
\mathrm{FT}\left[\mathrm{Skew}\right](\omega_k) = \left \Vert \sum_{j = 0}^{\mathcal{N}_t-1} f\left(t_j\right)e^{-i \omega_k t_j} \right \Vert, \,\,k = 0,...,\mathcal{N}_t -1,
\end{equation}
where $f(t) = \mathrm{Skew}(t) - \langle \mathrm{Skew}(t)\rangle_t$ and we subtract the average background signal $\langle \mathrm{Skew}(t)\rangle_t $ from the actual skewness $\mathrm{Skew}(t)$ to eliminate the peak at zero frequency. We define the discrete time as $t_j = j \Delta t$, where $\Delta t$ is the time step, and the total evolution time as $t_{\rm tot} = \mathcal{N}_t \Delta t$, with $\mathcal{N}_t$ the number of time steps. The discrete DC frequency components are given by $\omega_k = 2 \pi k/t_{\rm tot}$.

\begin{figure}[b!]
\includegraphics[width=0.98\columnwidth]{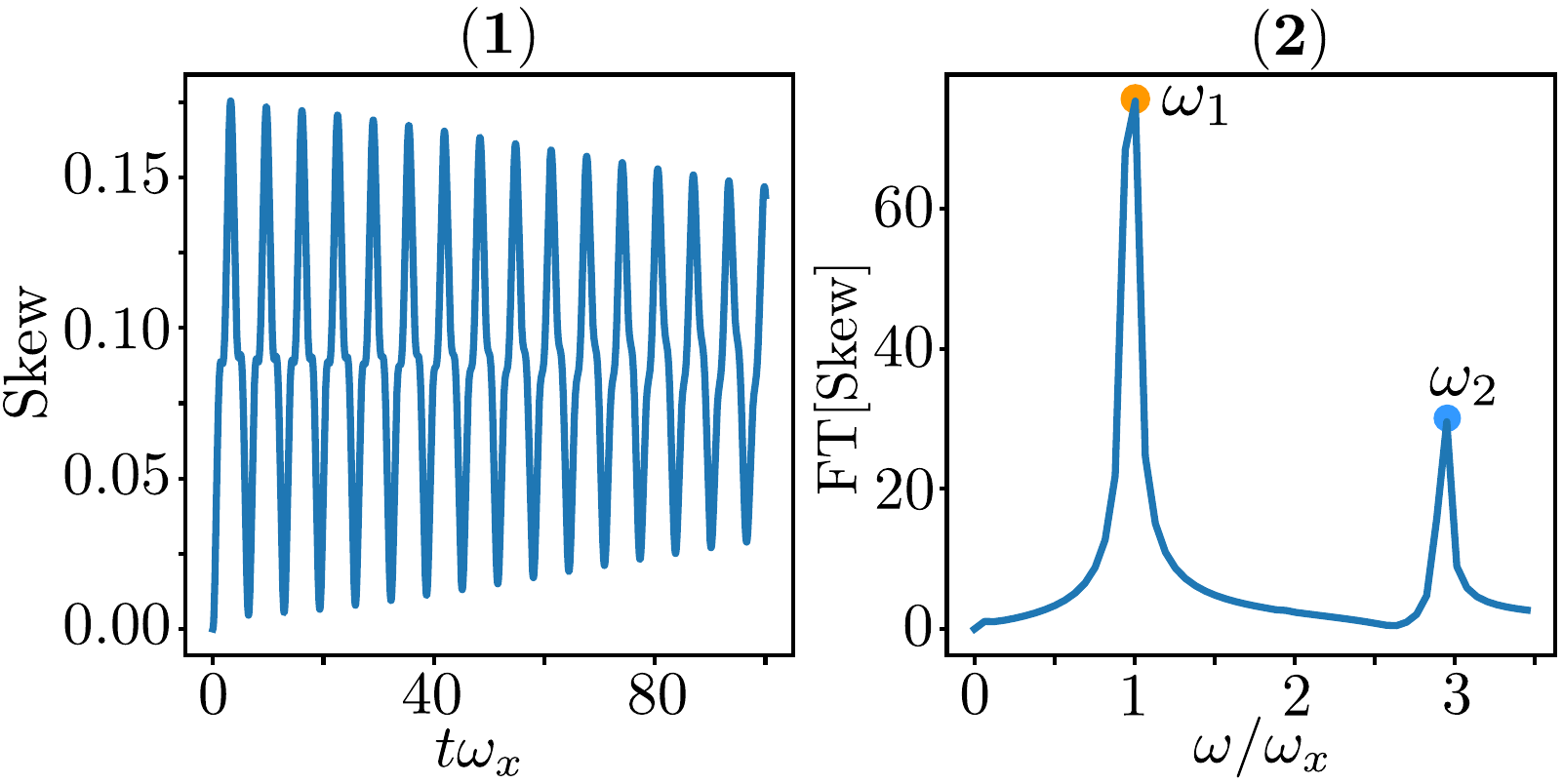}
\caption{~\textbf{(1)}:~Example of the skewness, Eq.~\eqref{Eq:Skew}, vs. the dimensionless time $t \omega_x$, calculated for dataset (c) (see Table \ref{Tab:Parameters}) at $\gamma_0\!=\!5.035$, corresponding to $N\!=\!1412$ in Fig.~\ref{Fig:N_gamma0}.~\textbf{(2)}:~Fourier transform of the skewness in \textbf{(1)}, $\rm {FT}\left[ \rm{Skew} \right]$, Eq.~\eqref{Eq:FT Skew}, vs. frequency $\omega/\omega_x$. The yellow and blue markers at the two peaks correspond to the DC oscillation frequencies $\omega_1$ and $\omega_2$.
}
\label{Fig:Skew}
\end{figure}

Figure~\ref{Fig:Skew} \textbf{(1)} shows a GHD calculation of the skewness $\mathrm{Skew}(t)$, Eq.~\eqref{Eq:Skew}, as a function of time.~Figure~\ref{Fig:Skew} \textbf{(2)} presents the corresponding Fourier transform, Eq.~\eqref{Eq:FT Skew}, where the positions of the peaks indicate the two dominant frequency components, $\omega_1$ and $\omega_2$.~Their relative excitation strengths are quantified by $K_1$ and $K_2 = 1-K_1$, where $K_1 = h_1^2/(h_1^2 + h_2^2)$ and $h_{1,2}$ are the peak heights \cite{Bayocboc2023}.~In this case, the DC oscillations exhibit a beating between $\omega_1 \simeq 1\omega_x$ and $\omega_2 \simeq 3\omega_x$, consistent with predictions for the collisionless regime \cite{DeRosi2016}, although with unequal strengths ($K_1>K_2$), unlike in Table \ref{Tab:DC freq}.

By repeating this analysis across datasets (a)--(e) in Fig.~\ref{Fig:Regimes_Diagram}, we extract the DC oscillation frequencies and their excitation strengths throughout the various regimes of the 1D Bose gas, as shown in Fig.~\ref{Fig:DC frequencies} of the main text.

\bibliography{Bibliography}

 \renewcommand{\theequation}{S\arabic{equation}}
 \setcounter{equation}{0}
 \renewcommand{\thefigure}{S\arabic{figure}}
 \setcounter{figure}{0}
 \renewcommand{\thesection}{S\arabic{section}}
 \setcounter{section}{0}
 \onecolumngrid

\end{document}


\flushbottom

\onecolumngrid
\newpage

\title{Supplemental Material for: ``Particle-hole origin of thermal beating in dipole-compression modes of a 1D Bose gas''}

\author{Caroline Mauron}
\affiliation{School of Mathematics and Physics, University of Queensland, Brisbane, Queensland 4072, Australia}

\author{Karen V. Kheruntsyan}
\email{karen.kheruntsyan@uq.edu.au}
\affiliation{School of Mathematics and Physics, University of Queensland, Brisbane, Queensland 4072, Australia}

\author{Giulia De Rosi}
\email{giulia.de.rosi@upc.edu}
\affiliation{Departament de F\'isica, Universitat Polit\`ecnica de Catalunya, Campus Nord B4-B5, 08034 Barcelona, Spain}

\maketitle

\twocolumngrid

\setcounter{equation}{0}
\setcounter{figure}{0}
\setcounter{table}{0}
\setcounter{page}{1}
\renewcommand{\theequation}{S\arabic{equation}}
\renewcommand{\thefigure}{S\arabic{figure}}
\renewcommand{\bibnumfmt}[1]{[S#1]}
\renewcommand{\citenumfont}[1]{S#1}

\section*{Generalized Hydrodynamics for the One-Dimensional Bose Gas}

The collective motion of a variety of \textit{non-integrable} many-body systems is often well described by classical hydrodynamics, which relies on only a few macroscopic conserved quantities—such as particle number, momentum, and energy—together with rapid thermalization leading to local thermodynamic equilibrium at high temperatures.  As a consequence of this rapid local thermalization,  at long times,  the system can be regarded as decomposed into slowly varying local “fluid cells", each described by a homogeneous Gibbs state.

By contrast,  generalized hydrodynamics (GHD) \cite{Castro-Alvaredo2016,Bertini2016} provides a reliable description of the time evolution of \textit{integrable} many-body systems, which admit exact solutions and possess infinitely many local conserved properties \cite{Thacker1981,Sutherland2004} that preclude thermalization \cite{Rigol2007, Rigol2008, cazalilla2010focus, Polkovnikov2011, Eisert2015, Gogolin2016}. By explicitly accounting for this infinite set of conserved quantities, GHD extends beyond classical hydrodynamics and is capable of capturing also the collisionless (non-hydrodynamic) regime.  The central assumption of GHD is that each local fluid cell of the system thermalizes to a generalized Gibbs ensemble (GGE),  whose parameters vary slowly in space and time.  For Bethe-ansatz-integrable models, such local GGEs are fully characterized by distributions of stable quasiparticle excitations. GHD then describes the large-scale dynamics as ballistic transport of these quasiparticles, with effective velocities self-consistently renormalized by interactions.~Beyond the two original works \cite{Castro-Alvaredo2016,Bertini2016} from 2016, there are now several review articles and additional insights on GHD available in the literature \cite{Doyon2017_PRL,Doyon2017,Schemmer2019,Doyon2020,Bastianello2021,Alba2021,Malvania2021,Kerr2023,Watson2023}. Here, we provide a summary of the formalism.

For the Lieb--Liniger model describing a one-dimensional Bose gas with repulsive contact interactions,  local macrostates are described within Yang-Yang's thermal Bethe ansatz \cite{Yang1969} by a quasiparticle rapidity $\lambda$ (in units of velocity, with $p = m\lambda$ being the quasimomentum) and the corresponding local (in space $x$ and time $t$) quasiparticle density $f_p(\lambda;x,t)$.~In the presence of a slowly varying external potential $V(x)$, integrability is weakly broken and GHD remains valid within the local density approximation, where a position-dependent chemical potential $\mu(x)=\mu_0-V(x)$ can be defined. In this case, the Euler-scale (first-order) GHD equation takes the form of a Bethe--Boltzmann convection equation \cite{Castro-Alvaredo2016,Bertini2016,Doyon2017}:
\begin{align}
\partial_t f_p(\lambda;x,t)
+ &\partial_x \!\left[ v^{\mathrm{eff}}[f_p]\, f_p(\lambda;x,t) \right]  \nonumber \\
& = \frac{\partial_x V(x)}{m}\, \partial_\lambda f_p(\lambda;x,t),
\label{eq:GHD_fp}
\end{align}
where $v^{\mathrm{eff}}[f_p]$ denotes the effective velocity of quasiparticle excitations expressed in terms of $f_p$. The right-hand side of Eq.~\eqref{eq:GHD_fp} accounts for the acceleration induced by the external force --arising from the potential $V(x)$-- while the left-hand side describes ballistic transport.

It is often convenient to introduce the occupation (filling) function,
\begin{equation}
\vartheta(\lambda;x,t)
=\frac{f_p(\lambda;x,t)}{f(\lambda;x,t)},
\end{equation}
where $f = f_p +f_h$ is the total density of the system's state and $f_h(\lambda;x,t)$ is the density of quasihole excitations.  Equation \eqref{eq:GHD_fp} can be rewritten in terms of $\vartheta(\lambda;x,t)$ as:
\begin{equation}
\partial_t \vartheta(\lambda;x,t)
+ v^{\mathrm{eff}}[\vartheta]\,\partial_x \vartheta(\lambda;x,t)
= \frac{\partial_x V(x)}{m}\,\partial_\lambda \vartheta(\lambda;x,t),
\label{eq:GHD_theta}
\end{equation}
which admits a simple interpretation in terms of characteristics in the phase space, i.e., the trajectories $x(t)$ along which quasiparticles move at their $v^{\mathrm{eff}}$.

The effective velocity $v^{\mathrm{eff}}(\lambda)$ incorporates many-body interaction effects via a dressing operation and is expressed in terms of the quasiparticle energy $E(\lambda)$ and momentum $p(\lambda)$ as \cite{Castro-Alvaredo2016,Bertini2016,Doyon2017}
\begin{equation}
v^{\mathrm{eff}}(\lambda)\equiv v^{\mathrm{eff}}[\vartheta(\lambda;x,t)]
= \frac{(\partial_\lambda E)^{\mathrm{dr}}}{(\partial_\lambda p)^{\mathrm{dr}}}.
\label{eq:v eff}
\end{equation}
For a generic bare function $g(\lambda)$, the dressing is defined by the linear integral equation
\begin{equation}
g^{\mathrm{dr}}(\lambda)
= g(\lambda) + \int\!\frac{d\lambda'}{2\pi}\,
\theta(\lambda-\lambda')\,\vartheta(\lambda')\,g^{\mathrm{dr}}(\lambda'),
\label{eq:dressing}
\end{equation}
where $\theta(\lambda)$ is the two-body scattering kernel, and the dependence on $x$ and $t$ variables has been suppressed for clarity.~For the Lieb--Liniger model, the kernel is
\begin{equation}
\theta(\lambda)=\frac{2\hbar g_{1\mathrm{D}}}{g_{1\mathrm{D}}^{\,2}+(\hbar\lambda)^2},
\end{equation}
where $g_{\rm 1D}$ is the interaction strength defined in the main text. 
The total density of the state is given by
\begin{equation}
f(\lambda)=f_p(\lambda)+f_h(\lambda)
= \frac{1}{2\pi m}\left(\partial_\lambda p\right)^{\mathrm{dr}} .
\end{equation}

Expectation values of conserved charge densities take a universal form within GHD \cite{Castro-Alvaredo2016,Bertini2016,Doyon2017}:
\begin{equation}
\langle q_i \rangle
= \frac{m}{\hbar} \int\! d\lambda\,
f_p(\lambda)\, h_i(\lambda)
= \frac{m}{\hbar} \int\!\frac{d\lambda}{2\pi}\,
\vartheta(\lambda)\, h_i^{\mathrm{dr}}(\lambda),
\label{eq:observables}
\end{equation}
where $h_i(\lambda)$ ($i=0,1,2,\ldots$) are the single-particle eigenvalues of the conserved quantities. In particular, $\langle q_0\rangle\equiv \langle q_0(x,t)\rangle=n(x,t)$ is the particle density (studied in the main text), $\langle q_1\rangle$ the momentum (mass current) density, and $\langle q_2\rangle$ the energy density.

In practice, our finite-temperature GHD simulations are performed by numerically solving Eq.~(\ref{eq:GHD_theta}) for the occupation function $\vartheta(\lambda;x,t)$ using the \emph{iFluid} framework \cite{Moller2020} as in Ref.~\cite{Watson2023}. The evolution is implemented via a discretization of the rapidity and spatial coordinates, combined with a characteristic-based time stepping for the convective equation. At each time step, the effective velocity $v^{\mathrm{eff}}(\lambda)$ \eqref{eq:v eff} is recomputed self-consistently from the instantaneous occupation function  $\vartheta(\lambda;x,t)$ through the dressing equation~(\ref{eq:dressing}).~Initial thermal states are obtained from local equilibrium Yang--Yang solutions within the local density approximation. The resulting dynamics captures the Euler-scale ballistic evolution of the system between locally equilibrated fluid cells.

Equations~(\ref{eq:GHD_fp})--(\ref{eq:GHD_theta}) thus constitute a closed Euler-scale hydrodynamic description of the one-dimensional Bose gas, sufficient for the analysis presented in the main text.~Corrections beyond the Euler scale, such as diffusive terms, are not required for the present work.

\bibliography{bibliography.bib}